\documentstyle[12pt]{article}
\begin{document}
\baselineskip=20pt
\textheight=21cm
\textwidth=17cm
\begin{center}
{\bf DEFORMATION QUANTIZATIONS WITH SEPARATION OF
VARIABLES ON A K\"AHLER MANIFOLD\\ A. V.  Karabegov}
\end{center}

{\bf Introduction}

In [Ka] a simple geometric construction
of some formal deformation quantization (see [BFFLS]) on a K\"ahler
manifold was introduced. This construction provides the deformation
quantization obtained from Berezin's
$*$-pro\-duct (see [Be])
on the orbits of a compact semisimple Lie group in [Mo2] and [CGR1] and
on bounded symmetric domains in  [Mo1] and [CGR2].

The formal $\star$-product on a K\"ahler manifold $M$ corresponding
to the quantization from [Ka] is connected with the separation of
variables into holomorphic and antiholomorphic ones in the
following sense. For each open subset $U\subset M$,
$\star$-multiplication from the left by a
holomorphic function and from the right by an antiholomorphic
function on $U$ coincides with the pointwise multiplication by
these functions.

It turns out that all such quantizations with separation of variables
can be obtained by a slightly generalized construction from [Ka]
and are completely parametrized by geometric objects, the formal
deformations of the original K\"ahler metrics.

The author wishes to take the opportunity to express his gratitude to\\
 B. V. Fedosov for stimulating discussions
and to J. H. Rawnsley for sending the preprints of his joint
papers with M. Cahen and S. Gutt.
\newpage

{\bf 1. Definition of deformation quantization with
separation of variables}

Define a formal deformation quantization on a symplectic manifold
 $M$ (see [BFFLS]).

Let $\{C_r(\cdot,\cdot)\},\ r=0,1,2,\dots$ be a family of
bidifferential operators on $M$, i.e., of differential operators which
act from $C^\infty(M)\otimes C^\infty(M)$ to $C^\infty(M)$.
Define a binary operation $\star$ in the space of formal power series
 ${\cal F}=C^\infty(M)[[\nu]]$, posing for
$f=\sum_{r=0}^\infty\nu^r f_r$ and $g=\sum_{r=0}^\infty\nu^r g_r$
\begin{equation}
f\star g=\sum_{r=0}^\infty \nu^r \sum_{i+j+k=r} C_i(f_j,g_k).
\end{equation}

The operation $\star$ defines a formal deformation quantization
on the symplectic manifold $M$, if it is associative and for
$f,g\in C^\infty(M)$ holds
\begin{equation}
C_0(f,g)=fg,\ C_1(f,g)-C_1(g,f)=i\{f,g\},
\end{equation}
where $\{\cdot,\cdot\}$ is a Poisson bracket on $M$,
corresponding to the symplectic structure.

In such a case the operation $\star$ is called
a $\star$-product.

All the deformation quantizations considered in this paper
are formal, so in the sequel we will not mention it explicitly.

Since a $\star$-product is given by differential operators,
it is local, that is, it can be restricted to any open subset $U\subset M$.
The restriction of $\star$ defines a $\star$-product
in the space ${\cal F}(U)=C^\infty(U)[[\nu]]$.

If there is given a deformation quantization on $M$
then for each open subset $U\subset M$ in the space ${\cal F}(U)$
act the algebras ${\cal L}(U)$ and ${\cal R}(U)$
of the left and right $\star$-multiplication operators,
respectively.
For $f,g\in{\cal F}(U)$ define the operators $L_f\in{\cal L}(U)$ and
$R_g\in{\cal R}(U)$ by the relations $L_fg=R_gf=f\star g$.

The operators from ${\cal L}(U)$ commute with the operators
from ${\cal R}(U)$, \\ $[L_f,R_g]=0$.

For $U=M$ denote ${\cal L}={\cal L}(M)$, ${\cal R}={\cal R}(M)$.

Let ${\cal D}(U)$ be the algebra of the formal series of differential
operators of the form $\tilde A=\sum_{r=0}^\infty \nu^r A_r$,
where $A_r$ are differential operators on $U$ with smooth coefficients.
These series act as linear operators on the space
${\cal F}(U)$, for
$\tilde A=\sum_{r=0}^\infty \nu^r A_r$ and
$f=\sum_{r=0}^\infty\nu^r f_r$
$$
\tilde Af=\sum_{r=0}^\infty \nu^r\sum_{s=0}^r A_{r-s}f_s.
$$
Since one can take a pointwise product of the elements of
${\cal F}(U)$, ${\cal F}(U)$ is included in ${\cal D}(U)$
as the algebra of pointwise multiplication operators.
It follows from the definition of $\star$-product that
${\cal L}(U)$ and ${\cal R}(U)$ are subalgebras of ${\cal D}(U)$.

Further, we will refer sometimes to formal series of functions,
operators etc., as to formal functions, operators, or even omit
the word formal, which must not lead to a misunderstanding.

Let  $M$ be a K\"ahler manifold of complex dimension $m$
with a K\"ahler form $\omega_0$
of the type $(1,1)$.

{\it Definition.} A deformation quantization on the
K\"ahler manifold $M$ is called a deformation quantization with
separation of variables if, for any open subset $U\subset M$
and functions $a,b,f\in C^\infty(U)$, such that $a$ is
holomorphic and $b$ antiholomorphic, holds
$a\star f=a\cdot f,\ f\star b=f\cdot b$.

If on $M$ there is defined a deformation quantization with
separation of variables, then for a holomorphic function $a$
and antiholomorphic function $b$ on an arbitrary open subset
$U\subset M$, the operators $L_a$ and $R_b$ are the operators
of pointwise multiplication by the functions $a$ and $b$
respectively, $L_a=a$ and $R_b=b$.  If, moreover, $U$
is a coordinate chart with holomorphic coordinates
$z^1,\dots,z^m$, then, since for  $f\in {\cal F}(U)$ the operator $L_f$
commutes with $R_{\bar z^l}=\bar z^l$, it contains only partial
derivatives by $z^k$.  Similarly, the operator
$R_f$ contains only partial derivatives by $\bar z^l$.

{\bf 2. Deformation of K\"ahler metrics corresponding to
quantization with separation of variables}

With each deformation quantization with separation of variables
on a K\"ahler manifold $M$ with a K\"ahler form $\omega_0$,
we canonically associate a formal deformation of the
K\"ahler metrics $\omega_0$, i.e., a formal series
$\omega=\omega_0+\nu\omega_1+\nu^2\omega_2+\dots$
such that $\omega_1,\omega_2,\dots$ are closed but not necessarily
nondegenerate forms of the type $(1,1)$ on $M$.

On a contractible coordinate chart $U$, there exists a
K\"ahler potential $\Phi_0\in C^\infty(U)$ such that
$\omega_0=i\partial\bar\partial\Phi_0=ig_{kl}dz^k\land d\bar z^l$,
where $g_{kl}=\partial^2\Phi_0/\partial z^k\partial\bar z^l$.
Here as well as below we use the tensor rule of summation over repeated
indices.
The K\"ahler potential $\Phi_0$ is defined up to a summand of
the form $a+b$, where $a$ is a holomorphic and  $b$
an antiholomorphic function on  $U$.

Denote by $(g^{lk})$ the inverse matrix to $(g_{kl})$.
The Poisson bracket of the functions $f,g\in C^\infty(U)$
can be expressed as follows:
$$
\{f,g\}=ig^{lk}(\frac{\partial f}{\partial z^k} \frac{\partial
g}{\partial \bar z^l}- \frac{\partial g}{\partial z^k}\frac{\partial
f}{\partial \bar z^l}).
$$

Let on $M$ be defined a deformation quantization. Introduce
bidifferential operators $D_r(\cdot,\cdot)$, such that for
$u,v\in C^\infty(M)$ holds $D_r(u,v)=C_r(u,v)-C_r(v,u)$. From
(2) it follows that $D_0=0$, and $D_1=i\{\cdot,\cdot\}$.
Thus for $f=\sum_{r=0}^\infty\nu^r f_r$ and
$g=\sum_{r=0}^\infty\nu^r g_r$ \begin{equation} f\star g-g\star
f=\sum_{r=1}^\infty \nu^r \sum_{i+j+k=r} D_i(f_j,g_k).  \end{equation}

{\it Lemma 1.} Let $U$ be a contractible coordinate chart on $M$.
The system of equations for an unknown function $u$ on $U$,
$D_1(u,z^k)=f^k,\ k=1,\dots,m$, where $f^k\in C^\infty(U)$,
has a solution if and only if for all $k,k'$ holds
$D_1(f^k,z^{k'})=D_1(f^{k'},z^k)$. Then the solution
$u$ is determined up to a holomorphic summand.

{\it Proof.} By using the fact that $D_1=i\{\cdot,\cdot\}$,
the lemma can easily be reduced to the assertion that the solvability
condition of the equation $\bar\partial u=g_{kl}f^kd\bar z^l$ is a
$\bar\partial$-closedness of the form $g_{kl}f^kd\bar z^l$.

{\it Proposition 1.} Let
on a K\"ahler manifold $M$ with a K\"ahler
form $\omega_0$ be defined
a formal deformation quantization with
separation of variables.
Then on each contractible coordinate chart
$U\subset M$ there exist formal functions
$u^1,\dots,u^m\in{\cal F}(U)$ such that
$u^k\star z^{k'}-z^{k'}\star u^k=\nu\delta^{kk'}$, where $\delta$
is the Kronecker symbol.

{\it Proof.} We will construct, say, the function $u=u^1$.
Let $u=u_0+\nu u_1+\nu^2 u_2+\dots$.  The coefficients $u_r$
have to satisfy the following system of equations,
\begin{equation} \sum_{r=0}^\infty \nu^r \sum_{s=0}^{r-1}
D_{r-s}(u_s,z^k)= \nu\delta^{1k},\ k=1,\dots, m.
\end{equation}
Equating the coefficients at the same powers of
$\nu$ on the left-hand and right-hand sides of (4)
we get, at $r=1$, the equations
$D_1(u_0,z^k)=\delta^{1k}, \ k=1,\dots, m$.  Taking into account
that $D_1=i\{\cdot,\cdot\}$,
it is easy to check that
the function $u_0=\partial\Phi_0/\partial z^1$
satisfies these equations.
For $r>1$, the obtained equations are as follows,
\begin{equation}
\sum_{s=0}^{r-1} D_{r-s}(u_s,z^k)=0,\
k=1,\dots, m.
\end{equation}

We construct the functions $u_s$ step by step
using equations (5) and lemma 1.  Assume that for $s<n$ the functions
$u_s$ are constructed and satisfy equations (5) for $r\leq n$.
We are going to show that the function $u_n$
can be found from equations (5) for
$r=n+1$, which we rewrite in the following form,
\begin{equation} D_1(u_n,z^k)=-\sum_{s=0}^{n-1}
D_{n-s+1}(u_s,z^k)=0,\ k=1,\dots, m.  \end{equation}
It follows from lemma 1 that equations (6) can be solved
for unknown $u_n$ if the sum $$ \sum_{s=0}^{n-1}
D_1(D_{n-s+1}(u_s,z^k),z^{k'}) $$ is symmetric with respect
to the permutation of the indices $k$ and $k'$.
 The Jacoby identity for the $\star$-commutator (3) is reduced to the
identities
\begin{equation} \sum_{i=1}^{r-1} D_i(D_{r-i}(f,g),h)+{\rm
cyclic\ permutation\ of}\ f,g,h=0 \end{equation} for any smooth
functions $f,g,h$.  Setting in (7)
$f=u_s,\ g=z^k,\ h=z^{k'},\ r=n-s+1$,
and taking into account that since $z^k$ pair-wise $\star$-commute,
$D_r(z^k,z^{k'})=0$ holds, we get \begin{equation}
\sum_{i=1}^{n+1-s}(D_i(D_{n-i-s+2}(u_s,z^k),z^{k'})-
D_i(D_{n-i-s+2}(u_s,z^{k'}),z^k))=0.
\end{equation}
Summing up equations (8) for $s=0,1,\dots,n-1$ and
changing the order of summation, we get
$$
\sum_{s=0}^{n-1}(D_1(D_{n-s+1}(u_s,z^k),z^{k'})-
D_1(D_{n-s+1}(u_s,z^{k'}),z^k))=
$$
\begin{equation}
-\sum_{i=2}^{n+1}\sum_{s=0}^{n-i+1}(D_i(D_{n-i-s+2}(u_s,z^k),z^{k'})-
D_i(D_{n-i-s+2}(u_s,z^{k'}),z^k)).
\end{equation}
It follows from the fact that $D_1(u_0,z^k)=\delta^{1k}\
\star$-commutes with $z^{k'}$ that\\
 $D_i(D_1(u_0,z^k),z^{k'})=0$, therefore the inner sum on the right-hand
side of (9) at $i=n+1$ is equal to zero. It follows from (5)
that the inner sum on the right-hand side of (9)
is equal to zero also for $1<i<n+1$,
thus the right-hand side of (9) equals zero
which proves the solvability of the system (6) for unknown $s_n$.
The proposition is proved.

In a completely analogous way, one can find the formal functions
$v^1,\dots,v^m$\\
$\in{\cal F}(U)$ such that $v^l\star
\bar z^{l'}-\bar z^{l'}\star v^l=\nu\delta^{ll'}$.

Since $L_{z^k}=z^k$, it follows from proposition 1 that
$[L_{u^k},z^{k'}]=\delta^{kk'}$.
Using the fact that the operators from ${\cal L}(U)$
contain only partial derivatives by $z^k$, the operators $L_{u^k}$
and similarly, operators $R_{v^l}$, can be calculated explicitly.

{\it Lemma 2}. $L_{u^k}=u^k+\nu\partial/\partial z^k$,
               $R_{v^l}=v^l+\nu\partial/\partial \bar z^l$.

Introduce the formal differential forms
$\alpha=-\sum_k u^kdz^k$
and $\beta=\sum_l v^ld\bar z^l$.
Since the operators $L_{u^k}$ and $R_{v^l}$ commute, one gets
$\partial u^k/\partial\bar z^l=\partial v^l/\partial z^k$, therefore
$\bar\partial\alpha=\partial\beta$.
Define the closed formal differential form
$\omega=i\bar\partial\alpha=i\partial\beta$ of the type $(1,1)$.
As follows from the proof of proposition 1, the first term of the
formal series $\omega$ coincides with
$\omega_0$, therefore $\omega$ is a deformation of the K\"ahler
form $\omega_0$.

Assume $\tilde
u^1,\dots,\tilde u^m$ is another set of solutions of (4), and set
$\tilde\alpha=-\sum_k \tilde u^kdz^k$.  It follows from lemma 2 and
from the fact that the operators
$L_{\tilde u^k}$ and $R_{v^l}$ commute, that the form
$i\bar\partial\tilde\alpha$ coincides with $\omega$, that is,
$\omega$ does not depend on the concrete choice of the solution of
system (4).  It is easy to show also that $\omega$ does not depend
on the choice of coordinates on $U$.

It follows from the Poincare $\bar\partial$-lemma that on a
contractible coordinate chart $U\subset M$ there exists a formal series
$\Phi=\Phi_0+\nu\Phi_1+\dots\in {\cal F}$ which is a potential
of the formal K\"ahler metrics
$\omega=\omega_0+\nu\omega_1+\nu^2\omega_2+\dots$. That means that
for all $r\geq 0,\quad$
$\omega_r=i\partial\bar\partial\Phi_r=
i(\partial\Phi^2_r/\partial z^k\partial\bar z^l)dz^k\land d\bar z^l$.

Since $\omega=i\bar\partial\alpha=i\bar\partial(-\partial\Phi)$,
then $\alpha+\partial\Phi$ is a $\bar\partial$-closed form of the
type $(1,0)$. Therefore the coefficients of
$\alpha+\partial\Phi$, which are equal to $\partial\Phi/\partial
z^k-u^k$, are holomorphic.  Now it is straightforward that
$L_{\partial\Phi/\partial z^k}= \partial\Phi/\partial
z^k+\nu\partial/\partial z^k$ and, similarly, $R_{\partial\Phi/\partial
\bar z^l}= \partial\Phi/\partial \bar z^l+\nu\partial/\partial\bar
z^l$.

Thus, starting from a given deformation quantization with separation
of variables, we construct on each contractible chart $U\subset M$
a formal deformation $\omega$ of the K\"ahler form $\omega_0$.
It follows from the construction of the form $\omega$
that on the intersections of charts the local forms agree with each
other and define a global form $\omega$ on $M$.

{\it Theorem 1.} Each deformation quantization with separation of
variables on a K\"ahler manifold $M$ canonically corresponds
to a formal K\"ahler metrics $\omega$, which is a
deformation of the K\"ahler metrics  $\omega_0$ on $M$.
  If $\Phi$ is a potential of the formal metrics
$\omega$ on a coordinate chart $U\subset M,\quad$ then\\
$L_{\partial\Phi/\partial z^k}= \partial\Phi/\partial
z^k+\nu\partial/\partial z^k$ and $R_{\partial\Phi/\partial \bar z^l}=
\partial\Phi/\partial \bar z^l+\nu\partial/\partial\bar z^l$.

{\bf 3. A construction of the quantization with separation of
variables from deformation of K\"ahler metrics}

Our goal is to generalize the construction of the deformation
quantization announced in [Ka].

Assume that there is given a formal deformation
$\omega$ of the K\"ahler metrics $\omega_0$ on $M$.

{\it Lemma 3.} Assume that on a contractible coordinate chart
$U\subset M$, there is chosen a potential
$\Phi=\Phi_0+\nu\Phi_1+\dots\in {\cal F}$ of the formal
metrics $\omega$.  Then the set of formal series of differential
operators from ${\cal D}(U)$, which commute with the
operators $\bar z^l$ and
$\partial\Phi/\partial \bar z^l+\nu\partial/\partial\bar z^l$,
depends only on the metrics $\omega$, rather than on the concrete
choice of the potential.

{\it Proof}. If $\Phi'\in{\cal F}$ is another potential of the
metrics $\omega$, then $\Phi'=\Phi+a+b$, where
$a$ and $b$ are formal series of holomorphic and antiholomorphic
functions, respectively.  An operator which commutes with $\bar z^l$
and $\partial\Phi/\partial \bar z^l+\nu\partial/\partial\bar z^l$,
commutes also with multiplication by antiholomorphic
functions. Therefore, it commutes with
$\partial\Phi'/\partial \bar z^l+\nu\partial/\partial\bar z^l=
(\partial\Phi/\partial \bar z^l+\nu\partial/\partial\bar z^l) +\partial
b/\partial \bar z^l$, which implies the assertion of the lemma.

Denote the set of formal operators mentioned in lemma 3 by
${\cal L}_\omega(U)$. Notice, that ${\cal L}_\omega(U)$
is an operator algebra.

Let $U$ be a coordinate chart on $M$
with a potential  $\Phi_0$ of the K\"ahler metrics
$\omega_0$ defined on it.  Denote by $S(U)$
the set of differential operators with smooth coefficients on
$U$, which commute with multiplication by the antiholomorphic
coordinates $\bar z^l$, i.e. which contain only partial derivatives
by $z^k$.

Define the differential operators $D^l$ on $U$,
$D^l=g^{lk}\partial/\partial z^k=i\{\bar z^l,\cdot\}$.

{\it Lemma 4.} For all $k,l,l'=1,\dots,m$ the following
relations hold\\ (i)  $[D^l,D^{l'}]=0$;\\ (ii)
$[D^l,\partial\Phi_0/\partial\bar z^{l'}]=\delta^l_{l'}$;\\
(iii)$\partial/\partial z^k=g_{kl}D^l$.

The assertion of the lemma can be checked by direct calculations.

It follows from lemma 4 that any operator from $S(U)$
can be canonically represented as a sum of monomials of the form
$a_{l_1\dots l_s}D^{l_1}\dots D^{l_s}$, where $a_{l_1\dots l_s}\in
C^\infty(U)$ is symmetric with respect to $l_i$.

{\it Definition.} The twisted symbol of an operator $A\in S(U)$,
which is represented in the canonical form
$A=\sum a_{l_1\dots l_s}D^{l_1}\dots D^{l_s}$,
is a polynomial in $\xi^1,\dots,\xi^m$,
$a(\xi)=\sum a_{l_1\dots l_s}\xi^{l_1}\dots
\xi^{l_s}$ with coefficients in $C^\infty(U)$.

{}From lemma 4 easily follows

{\it Lemma 5.} Let $a(\xi)$ be the twisted symbol of an operator
 $A\in S(U)$. Then the twisted symbol of the operator
 $[A,\partial\Phi_0/\partial\bar z^l]$ is equal to $\partial
a/\partial\xi^l$.

Consider a system of equations for an unknown operator
$A\in S(U)$,
\begin{equation}
[A,\partial\Phi_0/\partial\bar z^l]=B_l,\qquad l=1,\dots, m,
\end{equation}
where $B_l\in S(U)$.

{\it Lemma 6.} System (10) has solutions if and only if
for all $l,l'$\\
$[B_l,\partial\Phi_0/\partial\bar z^{l'}]=
[B_{l'},\partial\Phi_0/\partial\bar z^l]$.  If $A_0$ is a partial
solution of the system, then the general solution is of the form
$A_0+A_1$, where $A_1$ is an arbitrary multiplication operator.

{\it Proof.} Pass to the twisted symbols $a,b_l$ of the
operators $A,B_l$, respectively. System (10) transforms to the
equation $da=\sum_l b_ld\xi^l$, where $da=\sum_l
(\partial a/\partial\xi^l)d\xi^l$.
The assertion of the lemma is now reduced to a standard fact
concerning differential forms with polynomial coefficients,
which follows from Euler's identity.

{\it Proposition 2.}  Let $\omega_0$ be a K\"ahler metrics on
   $M$ and $U\subset M$ be a contractible coordinate chart.
  For each formal function
  $f=\sum \nu^r f_r\in{\cal F}(U)$
  there exists a unique formal series of differential
  operators  $\tilde A_f=\sum \nu^r A_r$ from
${\cal L}_{\omega_0}(U)$, such that $\tilde A_f1=f$. In particular,
  $A_0$ is a multiplication operator by the function $f_0$.

{\it Proof.} Since $\tilde A_f$ commutes with antiholomorphic functions,
all the operators $A_r$ are in $S(U)$. Let $\Phi_0$ be a potential
of the metrics $\omega_0$. The commutation condition of $\tilde A_f$
with
$\partial\Phi_0/\partial\bar z^l+\nu\partial/\partial\bar z^l$ is
equivalent to the system of equations
$[A_0,\partial\Phi_0/\partial\bar z^l]=0$ and
\begin{equation}
[A_r,\partial\Phi_0/\partial\bar z^l]= [\partial/\partial\bar
z^l,A_{r-1}].
\end{equation}
  Find all the terms of the series $\tilde A_f$ step by step.
It follows from lemma 5 that $A_0$ is a multiplication operator,
so $A_01=f_0$ implies that $A_0=f_0$. Assume that we have found
all the operators $A_r$ for $r<s$ which satisfy (11) and such that
$A_r1=f_r$.  Let us now show that $A_s$ can be found from
(11) for $r=s$, i.e., that the conditions of lemma 6 on the right-hand
side of (11) are satisfied,
$$
  [[\frac{\partial}{\partial\bar
z^l},A_{s-1}], \frac{\partial\Phi_0}{\partial\bar z^{l'}}]=
[[\frac{\partial}{\partial\bar z^{l'}},A_{s-1}],
\frac{\partial\Phi_0}{\partial\bar z^l}].
$$
It follows from the Jacoby identity for commutators that
$$
[[\frac{\partial}{\partial\bar
z^l},A_{s-1}],\frac{\partial\Phi_0}{\partial\bar
z^{l'}}]=
[[\frac{\partial}{\partial\bar z^l},
\frac{\partial\Phi_0}{\partial\bar z^{l'}}],A_{s-1}]+
[\frac{\partial}{\partial\bar z^l},
[A_{s-1},\frac{\partial\Phi_0}{\partial\bar z^{l'}}]]=
$$
$$
[\frac{\partial^2\Phi_0}{\partial z^l\partial\bar
z^{l'}}, A_{s-1}]+ [\frac{\partial}{\partial\bar
z^l},[\frac{\partial}{\partial\bar z^{l'}},A_{s-2}]].
$$
It is easy to check that the last expression is symmetric with respect
to the permutation of  $l$ and $l'$.  Thus system (11) is solvable for
$r=s$.
Among its solutions there is the only one solution
$A_s$ such that $A_s1=f_s$.
The assertion is proved.

{\it Lemma 7.} For a given formal function
$f=f_0+\nu f_1+\dots\in{\cal F}(U)$ there exists a function
$g=g_0+\nu g_1+\dots\in{\cal F}(U)$ such that $\tilde A_fg=1$
if and only if $f_0$ does not vanish on $U$. Then $g$ is defined
uniquely and $g_0=1/f_0$.

{\it Proof.} Let $\tilde A_f=\sum\nu^r A_r$.
The condition $\tilde A_fg=1$ is equivalent to the system
of equations
$A_0g_0=1$ and $A_0g_r=-\sum_{s=0}^{r-1}A_{r-s}g_s$.
According to proposition 2,  $A_0=f_0$, therefore if
$f_0$ does not vanish, all the functions $g_r$
can be calculated step by step. That completes the proof.

{\it Lemma 8.} Let the formal functions $f,g\in{\cal F}(U)$
be such that $\tilde A_fg=1$. Then the operator $\tilde A_g$
is inverse to $\tilde A_f$ and, in particular,
$\tilde A_gf=1$.

{\it Proof.} The operator $\tilde A_f\tilde A_g$
belongs to ${\cal L}_{\omega_0}(U)$.
Since $\tilde A_f\tilde A_g1=\tilde A_fg=1$, then
$\tilde A_f\tilde A_g=\tilde A_1=1$. It follows from lemma 7
that the coefficient at the zero power of
$\nu$ of the formal series $g$ does not vanish.
Therefore, there exists a function $h\in{\cal F}(U)$
such that $\tilde A_gh=1$, so $\tilde A_g\tilde A_h=1$.
Thus the operator $\tilde A_g$ has both left and right inverse
operators which immediately implies the assertion of the lemma.

We will use some elementary facts about formal series.
Let $R$ be a vector space and $\tilde R=R[[\nu]]$
be the space of formal series with coefficients in  $R$.
There is a decreasing filtration in  $\tilde R$,
$\tilde R=\tilde R_0\supset\tilde R_1\supset\tilde
R_2\dots$, where $\tilde R_n$ consists of the series of the form
$\tilde A=\sum_{r=n}^\infty \nu^r A_r$,  $A_r\in R$.
An element $\tilde A\in\tilde R$ is of the order $n,\ {\rm ord}(\tilde
A)=n$, if $\tilde A\in\tilde R_n\backslash\tilde R_{n+1}$.
A series $\sum\tilde A_n$ with the elements $\tilde A_n\in\tilde R$,
such that the order ${\rm ord}(\tilde A_n)\to\infty$ as $n\to\infty$,
converges to an element of $\tilde R$ with respect to the topology
defined by the filtration.
If $\tilde A-\tilde B$ is of the order $n$, we write
$\tilde A\equiv\tilde B\ ({\rm mod}\ \nu^n)$.

For an arbitrary formal function
$S=S_0+\nu S_1+\dots\in{\cal F}(U)$ define its exponent,
$e^S=e^{S_0}\sum_{n=0}^\infty(1/n!)(S-S_0)^n\in{\cal F}(U)$.
The series in the definition of the exponent converges since
${\rm ord}((S-S_0)^n)\geq n$.

{\it Lemma 9.} For $S\in{\cal F}(U),\quad\partial e^S/\partial z^k=
(\partial S/\partial z^k)e^S$ and $\partial e^S/\partial \bar z^l=
(\partial S/\partial \bar z^l)e^S$.
Moreover, for $S,T\in{\cal F}(U)$ holds the equality
$e^S\cdot e^T=e^{S+T}$.

The proof is standard.

{\it Proposition 3.}  Let $\omega$ be a formal deformation of
  the K\"ahler metrics  $\omega_0$ on $M$ and $U\subset M$
  be a contractible coordinate chart. For each formal function
  $g=\sum \nu^r g_r\in{\cal F}(U)$ there exists a unique
  formal series of differential operators
  $\tilde B_g=\sum \nu^r B_r$ from ${\cal L}_{\omega}(U)$ such that
  $\tilde B_g1=g$.

{\it Proof.}  Let $\Phi=\Phi_0+\nu\Phi_1+\Phi_2+\dots$ be a potential
of $\omega$. Set $S=\Phi_1+\nu\Phi_2+\dots$. It follows from lemma 9
that $e^{-S}(\partial\Phi_0/\partial\bar z^l+
\nu\partial/\partial\bar z^l)e^S=
\partial\Phi/\partial\bar z^l+\nu\partial/\partial\bar z^l$.
Since, moreover,  $e^{-S}\bar z^le^S=\bar z^l$
we get that $e^{-S}{\cal L}_{\omega_0}(U)e^S={\cal L}_\omega(U)$.
The operator $\tilde B_g$ exists if and only if there is a
function $f\in{\cal F}(U)$ such that $e^{-S}(\tilde A_f)e^S=\tilde
B_g$.  It is enough for $f$ to satisfy the relation
$e^{-S}\tilde A_f(e^S)=g$ or $\tilde A_f(e^S)=e^Sg$,
which is equivalent to the equality
$\tilde A_f\tilde A_{e^S}=\tilde A_{e^Sg}$.  From lemmas  7 and 8
it follows that there is a function $h\in{\cal F}(U)$ such that the
operator $\tilde A_h$ is inverse to $\tilde A_{e^S}$.
Therefore $\tilde A_f=\tilde A_{e^Sb}\tilde A_h$, and so
$f=\tilde A_{e^Sg}h$, which completes the proof.

According to proposition 3, the mapping $f\mapsto\tilde B_f$
is a bijection of ${\cal F}(U)$ onto ${\cal L}_{\omega}(U)$.
Since ${\cal L}_\omega(U)$ is an operator algebra,
one can define in ${\cal F}(U)$ an associative product
 $\star$, carrying over to ${\cal F}(U)$ the operator
product from ${\cal L}_{\omega}(U)$. For $f,g\in{\cal F}(U)$
by definition $\tilde B_{f\star g}=\tilde B_f\tilde B_g$.
Applying both sides of the obtained equality to the constant 1,
one gets $f\star g=\tilde B_fg$. That means that $\tilde B_f$
is a left multiplication operator in the algebra
${\cal F}(U)$ with the operation $\star$. Denote $L_f=\tilde B_f$.

Calculate the first two terms of the formal series of operators
$L_{\bar z^l}$.

{\it Lemma 10.} $L_{\bar z^l}\equiv \bar z^l+\nu D^l\ ({\rm mod}\
\nu^2)$.

{\it Proof.} Let $L_{\bar z^l} \equiv A+\nu B\ ({\rm mod}\
\nu^2)$, then

\begin{equation}
[L_{\bar z^l},\frac{\partial\Phi}{\partial \bar z^{l'}}]\equiv
[A+\nu B, \frac{\partial\Phi_0}{\partial \bar z^{l'}}+
\nu(\frac{\partial\Phi_1}{\partial \bar z^{l'}}+
 \frac{\partial}{\partial \bar z^{l'}})] \ ({\rm mod}\ \nu^2).
\end{equation}

The operators $L_{\bar z^l}$ and $\partial\Phi/\partial \bar z^{l'}$
commute, therefore the coefficients at the zero and first powers of
$\nu$ on the right-hand side of (12) are equal to zero.
First, $[A,\partial\Phi_0/\partial \bar z^{l'}]=0$, therefore,
according to lemma 5, $A$ is a multiplication operator.
Since $L_{\bar z^l}1=A1=\bar z^l$ then $A=\bar z^l$.
Taking into account that $A=\bar z^l$, we get the equation
$[B,\partial\Phi_0/\partial\bar z^{l'}]=\delta^l_{l'}$.
Since $B1=0$, from lemma 5 follows that $B=D^l$. The lemma is proved.

Now we obtain the formula expressing the operator
$L_f,\ f\in{\cal F}(U)$, via $L_{\bar z^l}$.

{\it Proposition 4.}
\begin{equation}
 L_f=\sum_\alpha \frac{1}{\alpha !}
(\frac{\partial}{\partial\bar z})^\alpha f\ (L_{\bar z}-\bar z)^\alpha,
\end{equation}
where $\alpha$ is a multi-index.

{\it Proof.} It follows from lemma 10 that
${\rm ord}(L_{\bar z^l}-\bar z^l)=1$, therefore\\
${\rm ord}((L_{\bar z}-\bar z)^\alpha)=|\alpha|$
so the series in (13) converges.
Denote temporarily the right-hand side of (13) by $\tilde A$.
Since $(L_{\bar z^l}-\bar z^l)1=0$, then $\tilde A1=f$,
so to prove the proposition it is enough to show that
$\tilde A\in{\cal L}_{\omega}(U)$.  Let $\alpha=(i_1,\dots,i_m)$
be a multi-index. Introduce the following notation,
 $\alpha\pm l=(i_1,\dots,i_l\pm 1,\dots,i_m)$.
 Taking into account that $L_{\bar z^l}\in{\cal L}_{\omega}(U)$, one
gets
 $$ [(\frac{\partial}{\partial\bar z})^\alpha f,
\frac{\partial\Phi}{\partial \bar z^l}+ \nu\frac{\partial}{\partial
\bar z^l}]=\nu (\frac{\partial}{\partial\bar z})^{\alpha+l}f\quad {\rm and}
$$ $$ [\frac{1}{\alpha !}(L_{\bar z}-\bar z)^\alpha,
\frac{\partial\Phi}{\partial \bar z^l}+
\nu\frac{\partial}{\partial \bar z^l}]=
-\nu\frac{1}{(\alpha-l)!}(L_{\bar z}-\bar z)^{\alpha-l},
$$
which implies that
$$
[\tilde A,\frac{\partial\Phi}{\partial \bar z^l}+
\nu\frac{\partial}{\partial \bar z^l}]=
\nu(\sum_\alpha \frac{1}{\alpha !}
(\frac{\partial}{\partial\bar z})^{\alpha+l}f\
(L_{\bar z}-\bar z)^\alpha-
 $$
 $$
\sum_\alpha \frac{1}{(\alpha-l)!}
(\frac{\partial}{\partial\bar z})^\alpha f\
(L_{\bar z}-\bar z)^{\alpha-l})=0.
$$
The proposition is proved.

It immediately follows from proposition 4 and bilinearity of
the product $\star$ that the product $\star$ is given by
formula (1) for some bidifferential operators $C_r$.
Let $u,v\in C^\infty(U)$.  Calculate the operators $C_0$ and $C_1$
considering the first two terms of the series $u\star v$
and taking into account lemma 10,
$$ u\star
v=L_uv\equiv uv+\nu\sum_l \frac{\partial u}{\partial\bar z^l} D^lv \
({\rm mod}\ \nu^2).
$$
It follows that $C_0(u,v)=uv$ and
$C_1(u,v)=\sum_l \partial u/\partial\bar z^l D^lv$, therefore
$$
C_1(u,v)-C_1(v,u)= \sum_l (\frac{\partial u}{\partial\bar z^l} D^lv-
\frac{\partial v}{\partial\bar z^l} D^lu)=
$$
$$ g^{lk}(\frac{\partial
v}{\partial z^k} \frac{\partial u}{\partial \bar z^l}- \frac{\partial
u}{\partial z^k}\frac{\partial v}{\partial \bar z^l})= i\{u,v\}.  $$
That means that the product $\star$ is a $\star$-product
on the chart $U$ with the K\"ahler metrics $\omega_0$.
It is clear from the construction of the product $\star$
from the deformation of K\"ahler metrics $\omega$ that on the
intersections of charts the products
$\star$ agree with each other and define a global
deformation quantization with separation of variables on the K\"ahler
manifold $M$. From theorem 1 it follows that the deformation of
K\"ahler metrics corresponding to the\\
 $\star$-product $\star$, coincides with $\omega$.  Thus we have stated
the following

{\it Theorem 2.} Deformation quantizations with separation of variables
on a K\"ahler manifold $M$ are in 1---1 correspondence with formal
deformations of the K\"ahler metrics $\omega_0$ on $M$.
If on $M$ there is given a quantization with separation of
variables corresponding to a formal deformation $\omega$ of
the metrics $\omega_0$, $U$ is a contractible coordinate chart
on $M$, and $\Phi$ is a potential of $\omega$ on $U$,
then the operators of left $\star$-multiplication ${\cal L}(U)$
are characterized by the property that they commute with multiplication
by antiholomorphic functions and with the operators
$R_{\partial\Phi/\partial \bar z^l}=
\partial\Phi/\partial \bar z^l+\nu\partial/\partial\bar z^l$.
Similarly, the operators of right $\star$-multiplication ${\cal R}(U)$
are characterized by the property that they commute with multiplication
by holomorphic functions and with the operators
$L_{\partial\Phi/\partial z^k}= \partial\Phi/\partial
z^k+\nu\partial/\partial z^k$.

\begin{center}
                  BIBLIOGRAPHY
\end{center}
{[BFFLS]}   F. Bayen, M. Flato, C. Fronsdal, A. Lichnerowicz and D.
Sternheimer, Deformation theory and quantization, {\it Ann.  Phys.}, v.
111, no.  1, 1978, p. 1-151.\\
{[Be]} F. A. Berezin, Quantization,
Math.  {\it USSR Izv.}, v.  8, 1974, p.  1109-1165.\\
{[CGR1]} M. Cahen, S.  Gutt, J.  Rawnsley, Quantization of
K\"ahler manifolds.  II, {\it Trans.  Amer.  Math. Soc.}, v.  337, no. 1,
1993, p. 73-98. \\
{[CGR2]}  M.  Cahen, S.  Gutt, J.  Rawnsley,
Quantization of K\"ahler manifolds. IV, to appear in Lett.  Math.
Phys.\\
{[Ka]} A. V. Karabegov, On deformation quantization on a
K\"ahler manifold, connected with Berezin's quantization, to appear in
Funct.  Anal. Appl.\\
{[Mo1]}   C. Moreno, $*$-products on some
K\"ahler manifolds, {\it Lett. Math.  Phys.}, v. 11, no.  4, 1986, p.
361-372.\\
{[Mo2]}  C. Moreno, Invariant star products and
representations of compact
semisimple Lie groups, {\it Lett.  Math. Phys.}, v.  12, no. 3, 1986, p.
217-229.
\end{document}